# Machine learning driven synthesis of few-layered WTe$_2$

*Manzhang Xu[1,3,‡], Bijun Tang[3,‡], Chao Zhu[3], Yuhao Lu[4], Chao Zhu[3], Lu Zheng[2], Jingyu Zhang[3], Nannan Han[2], Yuxi Guo[1,3], Jun Di[3], Pin Song[3], Yongmin He[3], Lixing Kang[3], Zhiyong Zhang[1]\*, Wu Zhao[1], Cuntai Guan[4], Xuewen Wang[2]\*, Zheng Liu[3,5]\**

[1] School of Information Science and Technology, Northwest University, Xi'an, Shannxi 710127, P. R. China.

[2] Institute of Flexible Electronics, Northwestern Polytechnical University, 127 West Youyi Road, Xi'an, 710072, P. R. China.

[3] School of Materials Science and Engineering, Nanyang Technological University, Singapore 639798, Singapore.

[4] School of Computer Science and Engineering, Nanyang Technological University, Singapore 639798, Singapore.

[5] CINTRA CNRS/NTU/THALES, UMI 3288, Research Techno Plaza, 50 Nanyang Drive, Border X Block, Level 6, Singapore 637553, Singapore.

[‡] These authors contributed equally to this work.

\* Corresponding author: zhangzy@nwu.edu.cn, iamxwwang@nwpu.edu.cn, z.liu@ntu.edu.sg.

**Reducing the lateral scale of two-dimensional (2D) materials to one-dimensional (1D) has attracted substantial research interest not only to achieve competitive electronic device applications but also for the exploration of fundamental physical properties. Controllable synthesis of high-quality 1D nanoribbons (NRs) is thus highly desirable and essential for the further study. Traditional exploration of the optimal synthesis conditions of novel materials is based on the trial-and-error approach, which is time consuming, costly and laborious. Recently, machine learning (ML) has demonstrated promising capability in guiding material synthesis through effectively learning from the past data and then making recommendations. Here, we report the implementation of supervised ML for the**

chemical vapor deposition (CVD) synthesis of high-quality 1D few-layered $WTe_2$ nanoribbons (NRs). The synthesis parameters of the $WTe_2$ NRs are optimized by the trained ML model. On top of that, the growth mechanism of as-synthesized 1T' few-layered $WTe_2$ NRs is further proposed, which may inspire the growth strategies for other 1D nanostructures. Our findings suggest that ML is a powerful and efficient approach to aid the synthesis of 1D nanostructures, opening up new opportunities for intelligent material development.

Machine learning (ML) has showed enormous promise in numerous research fields such as speaker recognition, image recognition, chemistry, genetics and genomics, biology, physics, as well as materials science[1-5]. Recently, ML has demonstrated great capability to accelerate the design and discovery of novel materials, through effectively learning from past and even failed data [6-17]. For example, Oliynyk et al.[14] predicted the crystal structure of equiatomic ternary compositions solely based on the constituent elements using ML model, which can correctly identify the low-temperature polymorph from its high-temperature form with high confidence. Lu et al.[15] discovered new hybrid organic-inorganic perovskites (HOIPs) for photovoltaic applications with ML, and six orthorhombic lead-free HOIPs with proper band gap and thermal stability were identified from 5158 unexplored HOIPs.

So far ML has been primarily employed for the discovery of novel materials and prediction of material properties[14-17]. ML-driven synthesis remains less studied, mainly due to the highly complicated synthesis process and small dataset. Successful synthesis of novel materials is critical towards its future application, whereas substantial effort is generally required with tradition laboratory exploration[18, 19]. Therefore, the feasibility and potential of introducing ML into the materials synthesis to expedite the exploration period and reduce the cost, is worthy to be investigated.

From the materials' aspect, one-dimensional (1D) nanostructures derived from two-dimensional (2D) transition metal dichalcogenides (TMDs) have recently been spotlighted, attributed to the intriguing properties arise from the lower dimensionality[20-24]. In the past decades, 2D TMDs have demonstrated great performance in the nanoelectronic and spintronic applications, and showed great potential for the fabrication of quantum devices[21-25]. Considering that the quantum confinement effect may become more prominent with the reduction of dimensionality, it is of great research interest to reduce the lateral scale of 2D TMDs to 1D. Among TMDs, tungsten ditelluride ($WTe_2$), with rich structural variations and strong spin-orbit coupling, is one of the most promising type-II Weyl semimetal and topological material. It provides an excellent platform for the investigation of quantum physical phenomena such as spintronics, superconductivity, quantum spin Hall Effect and Majorana fermion[26-29]. Even though chemical vapor deposition (CVD) has been widely adopted to grow high-quality 2D materials, controllable synthesis of 1D $WTe_2$ with CVD still remains challenging, due to the low activity of Te, undesired by-product formed between Te and $SiO_2$/Si substrate, small electronegativity difference between Te and W, as well as the poor stability of telluride[23, 30]. Therefore, effective identification of optimal synthesis conditions of 1D $WTe_2$ is urgently required.

In this work, we implement supervised ML on the CVD synthesis of few-layered $WTe_2$ and the paradigm is schematically depicted in Fig. 1. To optimize the synthesis condition, CVD synthesis data are collected experimentally to train the ML model. The well-trained model is capable to predict the probability of successful synthesis given a set of CVD parameters and then recommend the most favorable conditions. Information such as feature importance is further extracted from the trained model, providing new insights into the CVD synthesis of $WTe_2$. In addition, the growth mechanism of 1D $WTe_2$ NRs is further proposed, which suggests that $H_2$ gas flow rate governs the transformation from 2D to 1D, and the source ratio (Te/W) dominates the length-width ratio of $WTe_2$ NRs.

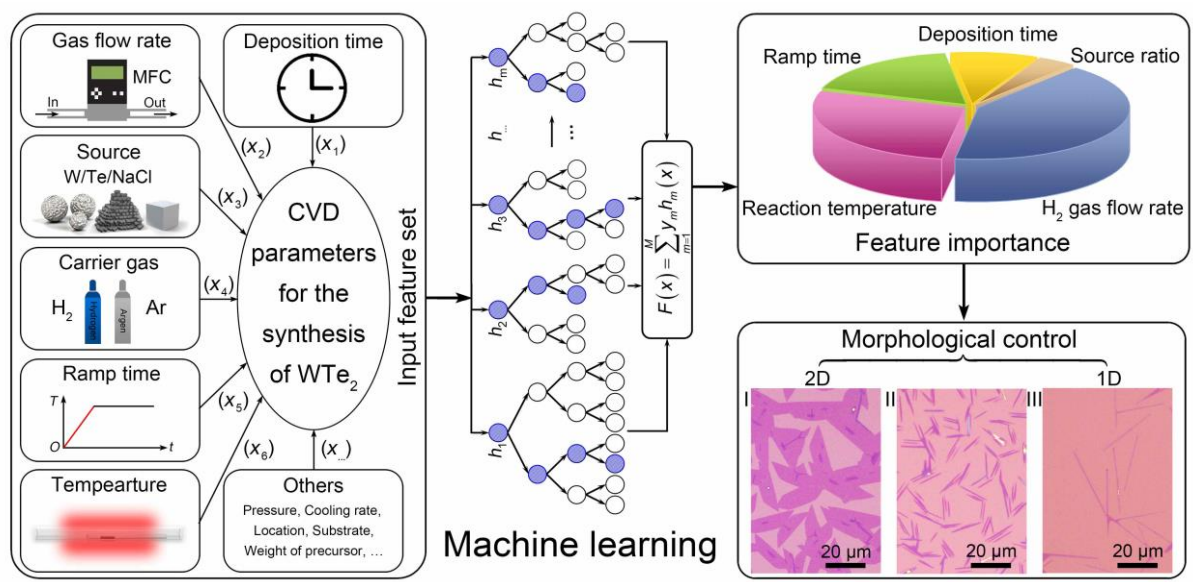

**Fig. 1** Design framework for the realization of morphology-controlled CVD synthesis of $WTe_2$ based on machine learning and experiments.

**Results**

**Identification of the optimal CVD synthesis ranges of $WTe_2$ by ML.** The CVD setup for the synthesis of $WTe_2$ NRs is shown in Fig. 2a. Empirically, five CVD parameters were identified as significant input features: source ratio (Te/W molar ratio, the molar ratio of tellurium power and tungsten power), reaction temperature (in °C), ramp time (in min), deposition time (in min) and $H_2$ gas flow rate (in sccm). 255 experiments were carried out in the lab with different combinations of growth parameters. Among them, $WTe_2$ were successfully synthesized in 141 experiments, whereas the rest 114 experiments showed negative result (no $WTe_2$ can be observed), as identified by the optical microscopy and Raman spectroscopy. A binary classification model is thus developed by defining "Can grow" as positive class and "Cannot grow" as negative class. XGBoost classifier (XGBoost-C) together with 10-fold nested cross validation have been applied in this study, as reported in our previous work[18]. Excellent performance of as-selected model has been demonstrated through plotting the receiver operating characteristic (ROC) curve, as shown in Fig. 2b. Area under ROC (AUROC) is commonly used to evaluate the ability of the model to successfully differentiate between two classes. AUROC of 1.0 represents perfect differentiation, whereas an area of 0.5 (indicated by the dashed line) is equivalent to a random guess. As displayed, our trained model possesses high AUROC of 0.93, indicating great discriminating capability between "Can grow" and "Cannot grow" classes of the trained model.

In order to further understand the system of CVD-grown $WTe_2$, feature importance has been extracted from the trained XGBoost-C model, with result shown in Fig. 2c. As indicated, $H_2$ gas flow rate is the most important feature affecting the synthesis of $WTe_2$, followed by the reaction temperature and ramp time, whereas source ratio is the least significant. Due to the difficulty of achieving controllable synthesis of the intriguing few-layered $WTe_2$, optimization of the synthesis condition has been carried out as per our previous study[18]. Based on the input ranges of each feature for optimization (Supplementary Table 1), respective "Can grow" probabilities (P) of the resulted 720,720 combinations have been predicted. The correlation of the output with each feature has been individually examined, with results shown in Fig. 2d. It can be seen clearly that the values of $H_2$ gas flow rate, reaction temperature and ramp time have the dominant effect on the distribution of P. This is consistent with the feature importance result discussed above (Fig. 2c). Among all the 720,720 synthesis conditions, 8 combinations with the highest P have been chosen for experimental verification. Respective results of the experiments have been summarized in Supplementary Table 2, which all demonstrate positive results. This suggests that controllable synthesis of few-layered $WTe_2$ can be realized with our trained ML model. In order to better interpret the information embedded in the as-predicted 720,720 conditions, we have further defined "Can grow" ratio. P of 0.5 (as indicated by the red dotted line in Fig. 2d) is adopted as the threshold where P > 0.5 represents "Can grow". Thus, the "Can grow" ratio for each feature at different values is defined as the number of "Can grow" conditions at that value over 720,720. The "Can grow" ratio for each feature at different values is illustrated in Fig. 2e. In this work, the range with "Can grow" ratio larger than 90% is selected as the optimized parameter range. As a result, with the aid of ML, the optimal ranges of synthesis parameters (see Supplementary Table 3) for $WTe_2$ with high P are determined as follows: $H_2$ gas flow rate (25 to 50 sccm), reaction temperature (600 to 750 ˚C), ramp time (18 to 20 min), deposition time (4 to 7 min), and source ratio (1 to 2).

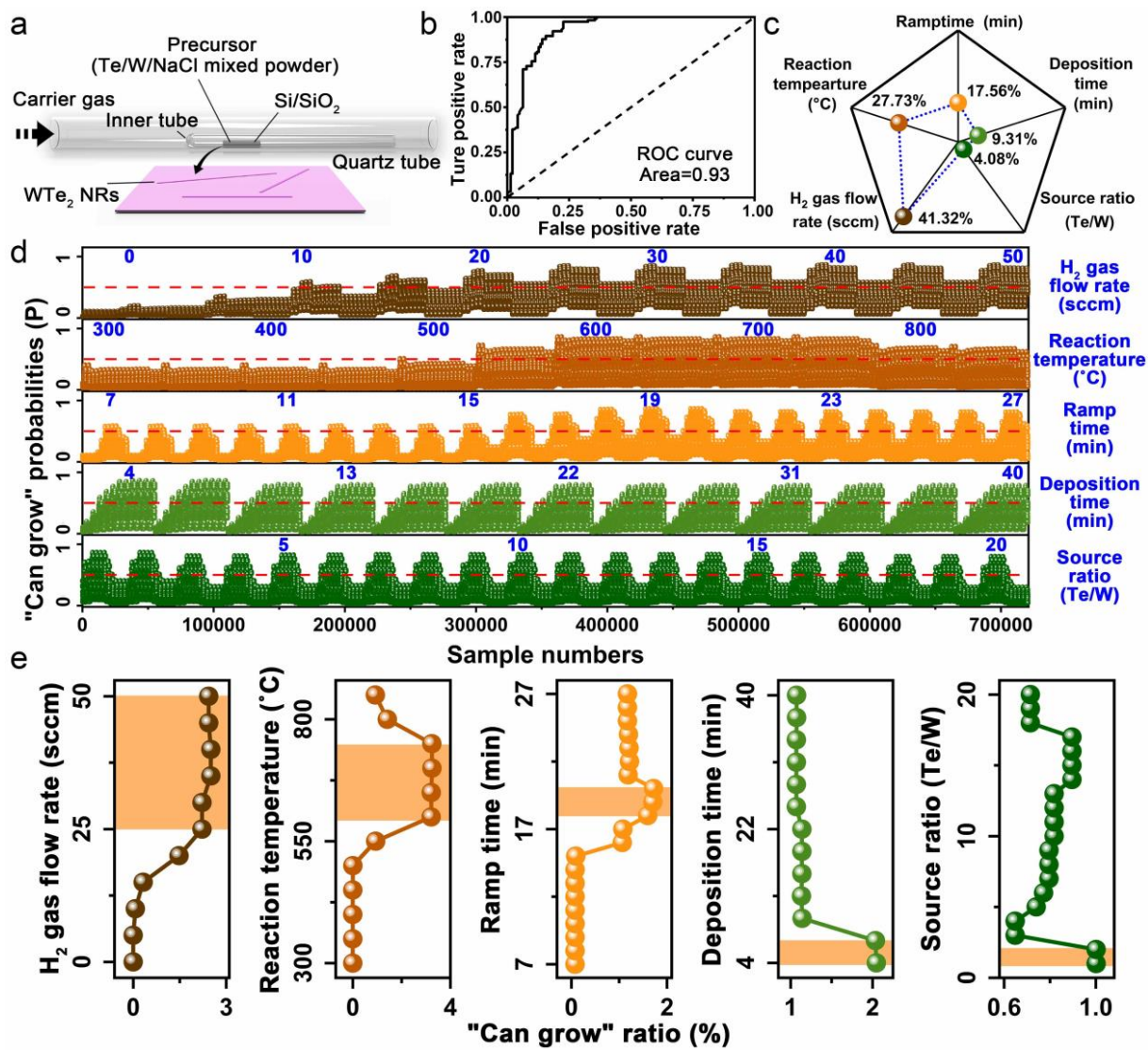

**Fig. 2** CVD synthesis of WTe$_2$ NRs and intelligent analysis with ML. **a** Schematic of the CVD setup used for growing WTe$_2$ NRs. **b** ROC curve for XGBoost-C model. **c** Feature importance of five synthesis parameters extracted from the well-trained XGBoost-C model. **d** Predicted "Can grow" probability (P) of all the 720,720 conditions with respect to each feature. P = 0.5 is represented by the red dotted line. **e** Plot of "Can grow" ratio for each feature at different values.

**Source ratio-dependent morphology evolution of WTe$_2$.** Based on the feature importance extracted from the trained ML model, source ratio (Te/W) is determined to be the least significant parameter affecting the "Can grow" probability of WTe$_2$. However, through experiments, the source ratio is found to be a critical factor governing the morphology of formed WTe$_2$ NRs. A series of experiments have been carried out with various source ratio, while the other four parameters are set within the optimal ranges as discussed above, to ensure the samples can be successfully synthesized. Specifically, reaction temperature, ramp time, deposition time and carrier gas flow rate are fixed at 750 ˚C, 20 min, 4 min, 100 sccm Ar and 40 sccm H$_2$, respectively. In order to systematically study the size evolution of WTe$_2$ NRs along with the change of source ratio, the length and width of the NRs have been examined based on the optical images. The frequency distribution histograms of the length-width ratio, width, and length with corresponding optical images at different source ratio (Te/W) are presented in Fig. 3a-c and Supplementary Fig. 1a. All the frequency distribution plots can be well fitted by the gamma distribution fitting curve. In order to further investigate the relationship between the source ratio and the size of as-obtained NRs, the mean value of length-width ratio, length, and width are calculated together with the corresponding standard deviation (Fig. 3d and Supplementary Fig. 1b). The length-width ratio shows a consistent increase from 12.5 to 77.7 as the source ratio increases from 1 to 20. Such intriguing geometrical evolution might be attributed to the concentration of Te, which will be discussed in detail later. Within the error margin, the length of WTe$_2$ NRs increases from 15.4 to 49.2 μm (Supplementary Fig. 1b) along with the increase of source ratio while the width shows the opposite trend (from 1.9 to 0.7 μm, Fig. 3d). Although the highest length-width ratio can be obtained at the source ratio of 20, considering width itself is an important attribute of NRs, source ratio of 16 is determined to be optimal in terms of both high length-width ratio and small width. As a result, the optimal synthesis condition of few-layered WTe$_2$ NRs has been determined with high P and small width (as described in the experiment section).

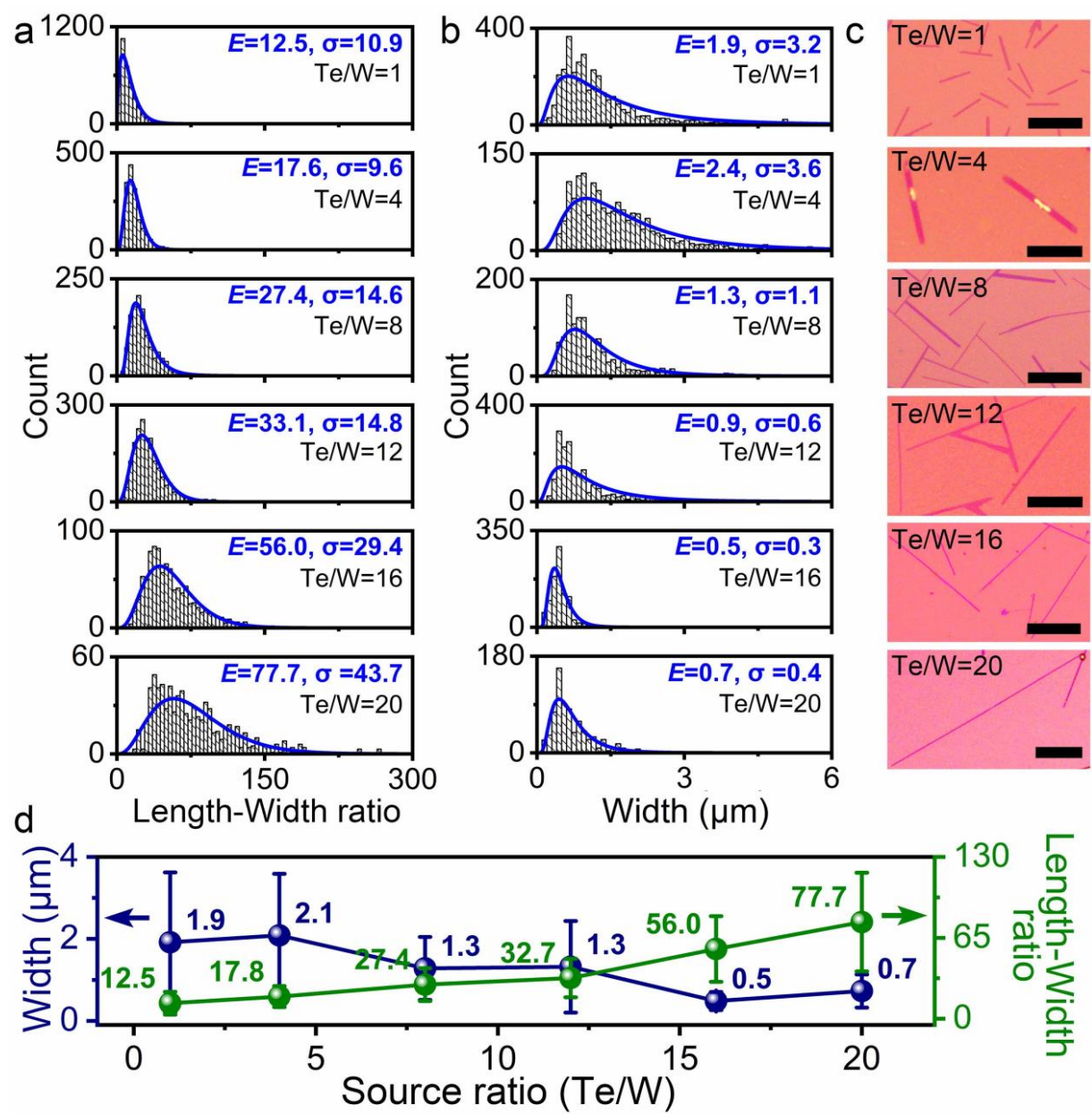

**Fig. 3** Source ratio-dependent morphology evolution of WTe$_2$. **a, b** The frequency distribution histograms of length-width ratio and width of WTe$_2$ NRs under different Te/W molar ratio and corresponding gamma distribution fitting curve (with the mean value ($E$) and standard deviation ($\sigma$)). **c** The optical images of WTe$_2$ NRs obtained under different source ratio (scale

bar, 20 μm). **d** Plot of the mean values (*E*) of width and length-width ratio with standard deviations (*σ*) versus Te/W molar ratio.

**Characterization of few-layered 1T' WTe₂ NRs.** The single crystalline few-layered $WTe_2$ NRs have been controllably synthesized under the optimal synthesis condition, with results shown in Fig. 4. The top and side views of the atomic structure of 1T' $WTe_2$ NRs are presented in Fig. 4a. Figure 4b shows the typical optical image of the single crystalline few-layered $WTe_2$ NRs with a length of ~50 μm. The width and thickness were further determined by the atomic force microscopy (AFM) with results shown in Fig. 4c. The width of the $WTe_2$ NRs is ~0.8 μm, and the thickness is ~2.5 nm indicating its triple-layer nature. All the $WTe_2$ possesses a ribbon-like anisotropic morphology (see Supplementary Fig. 2) with flat surface and uniform thickness. Raman spectroscopy was carried out to investigate the quality of as-synthesized single crystalline $WTe_2$ NRs, with the Raman spectrum displayed in Fig. 4d. Six vibrational modes of single crystalline $WTe_2$ NRs can be clearly observed, $B_1^{10}$, $A_2^4$, $A_2^3$, $A_1^4$, $A_1^7$, $A_1^9$ at 87.3, 108.6, 113.7, 132.4, 161.1, 210.6 cm⁻¹, respectively, which are in good agreement with previously reported values, suggesting high-quality of our synthesized WTe₂ NRs[31, 32]. To further demonstrate the quality and uniformity of $WTe_2$ NRs, the optical image and corresponding Raman intensity mapping ($A_1^9$) are shown in Fig. 4e. The Raman mapping region is highlighted by the red box in the optical image with a size of 6 μm × 60 μm. The uniform color distribution of the mapping region indicates the high homogeneity and low defect of the synthesized 1T′ $WTe_2$ NRs. X-ray photoelectron spectroscopy (XPS) was carried out to examine the chemical composition of the as-synthesized $WTe_2$ NRs (Supplementary Fig. 3).

The peaks of Te, W, Si, O, and C can be clearly observed in the spectrum with no extra peaks. The high-resolution XPS spectra of Te $3d$ and W $4d$ are shown in Fig. 4f. The Te $3d_{3/2}$ and $3d_{5/2}$ peaks, located at 583.7 and 573.3 eV, respectively, are associated with W-Te bonds[31, 33]. While the prominent W $4d$ peaks located at 256.9 and 243.6 eV, corresponding to $4d_{3/2}$ and $4d_{5/2}$, respectively[31, 33]. The chemical composition was calculated to be $WTe_{2.08}$ from the XPS results, indicating our CVD-grown $WTe_2$ NRs are considerably stoichiometric. Scanning transmission electron microscopy (STEM) images reveal the atomic structure of as-synthesized $WTe_2$ NRs. The typical bright-field TEM image of $WTe_2$ NRs is shown in Fig. 4g. The STEM image in Fig. 4h indicates that the 1T' $WTe_2$ consisting of 1D zigzagging W-Te chains along the axis of the unit cell, belonging to space group *$P2_1/m$ (Pmn2$_1$)*. EDX mappings in Fig. 4i verify the uniform distribution of elements. All the results indicate that the high-quality few-layered $WTe_2$ NRs have been successfully synthesized by our CVD method.

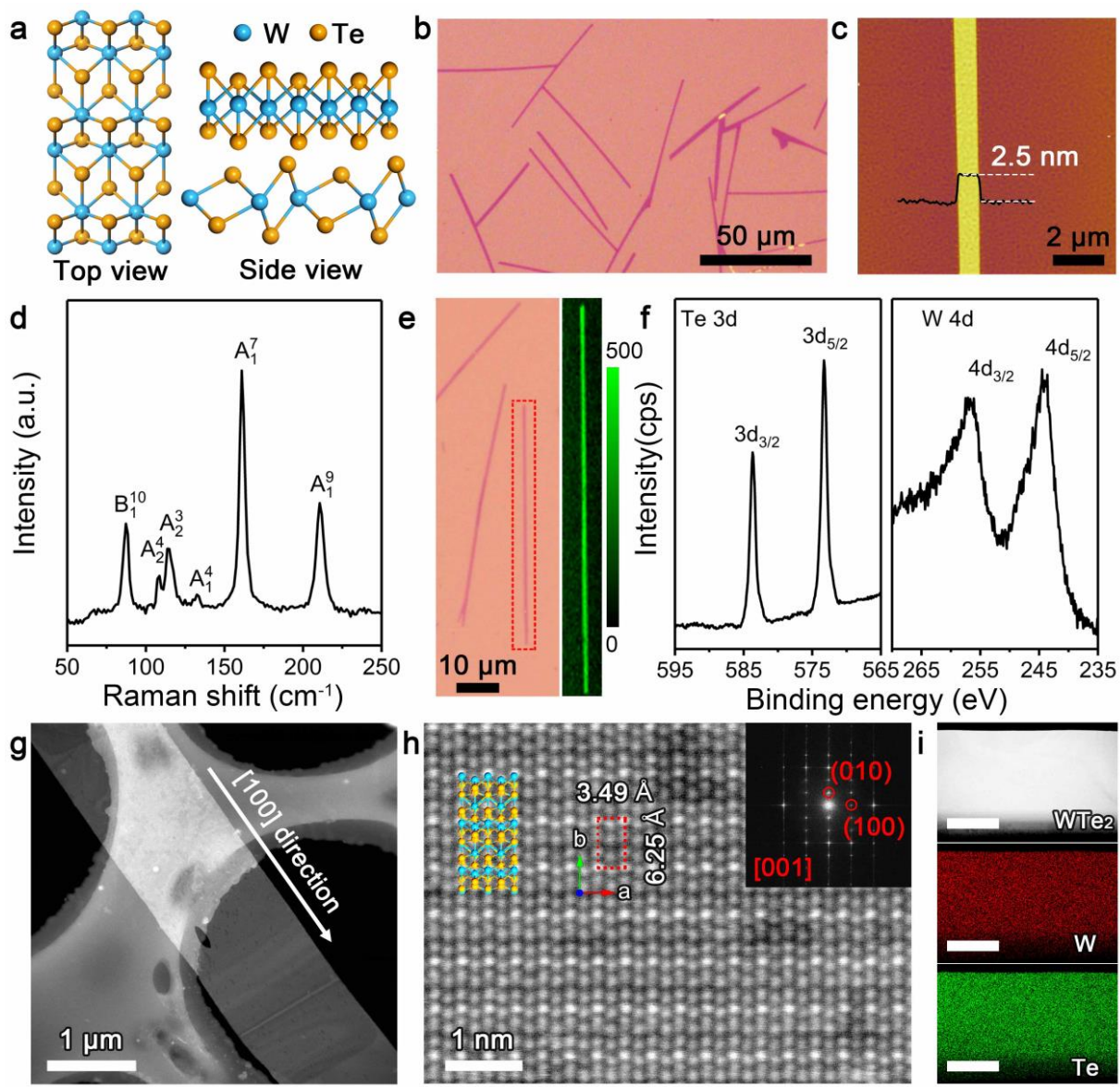

**Fig. 4** Characterization of as-synthesized few-layered WTe$_2$ NRs. **a** Top and side views of the crystal structure of 1T′ WTe$_2$. **b** Optical image of single crystalline WTe$_2$ NRs. **c** AFM result of WTe$_2$ NRs. **d, e** Raman spectra and mapping of WTe$_2$ NRs. **f** XPS spectra of WTe$_2$ NRs of Te 3d peaks and W 4d peaks. **g - i** Bright-field TEM image, STEM image (Inset, FFT pattern) and EDX mapping of WTe$_2$ NRs (scale bar, 1 μm).

**Growth mechanism governing the geometrical transformation from 2D to 1D.** Considering the high melting point and low reactivity of W, the mixed powder of NaCl/Te/W is adopted as the precursor and placed into the inner tube for the growth of few-layered WTe$_2$ NRs under H$_2$ atmosphere (see Supplementary Table 4 and Supplementary Note 1 for more information). Usually, deposition time of only a few minutes is required for the CVD synthesis of monolayer or few-layered TMDs as the synthesis process is quite fast[31, 32, 34]. In order to clearly observe the morphological evolution, synthesis conditions with high P from the ML results have been chosen to study the growth mechanism of WTe$_2$ NRs (H$_2$ gas flow rate of 30 sccm, reaction temperature of 650 ˚C, ramp time of 18 min, source ratio (Te/W) of 16, reaction time of 5 min). Here, the reaction time of H$_2$ is the only variable studied. Three controlled experiments have been carried out: (i) The H$_2$ gas was only introduced into the system when the temperature reaches 650 ˚C and closed after 5 min. (ii) The H$_2$ gas was introduced at room temperature and closed after 5-min deposition at 650 ˚C. (iii) The H$_2$ gas was kept throughout the whole process. Corresponding optical images of as-synthesized few-layered WTe$_2$ are shown in the inset I-III of Fig. 1, respectively. The morphology of as-formed few-layer WTe$_2$ evolved from 2D to 1D can be cleared observed. Based on the experimental results, a growth model for WTe$_2$ NRs has been further proposed (Fig. 5a-c). The synthesis of WTe$_2$ NRs could be divided in to three stages: formation of 2D WTe$_2$, H$_2$ etching of 2D WTe$_2$ and formation of WTe$_2$ NRs, the ripening regrowth of WTe$_2$ NRs.

The following reaction mechanism is proposed to explain the formation of the WTe$_2$: for the CVD growth of layered WTe$_2$ NRs with the Te/W/NaCl mixed powder under the Ar/H$_2$ atmospheric pressure (Fig. 5a).

$$Te(l) + H_2(g) \leftrightarrow H_2Te(g) \qquad (1)$$

$$W(s) + H_2Te(g) \leftrightarrow WTe_2(s) + H_2(g) \qquad (2)$$

The Gibbs free energy of as-proposed conversion reaction has been further calculated to investigate the role played by $H_2Te$ to reduce the barrier height (Fig. 5e and Supplementary Note 2). As calculated, with the aid of $H_2$, the synthesis of $WTe_2$ from W and the intermediate product $H_2Te$ requires ~135 kJ/mol less energy than the reaction between W and Te directly without $H_2$. In this regard, $H_2Te$ serves as an efficient Te carrier as well as a catalyst. Under Te-rich condition, the obtained $H_2$ in Reaction 2 could then react to form $H_2Te$ (Reaction 1), ensuring continuous tellurization process. Liquid W in the W-Te system then reacts with $H_2Te$, and the $WTe_2$ layered materials can be formed. All the reactions are reversible, and the direction is dependent on the concentration of the source and reaction temperature. Therefore, synthesis and decomposition of $WTe_2$ can occur simultaneously. In the first stage, the growth process is much faster than the decomposition of $WTe_2$ owing to the sufficient source supply, and $WTe_2$ can be synthesized quickly. As the chemical reaction proceeds, the growth speed starts dropping due to the reduced source concentration, and the $H_2$ etching of layered $WTe_2$ becomes significant. And it has been reported that excessive $H_2$ can decompose TMDs at high temperature[35, 36]. The etching will prefer [100] direction of 1T' $WTe_2$ structure as a result of the different bond lengths of W-Te, demonstrated in Fig. 5d. As the weak W-Te bonds can be easily broken by $H_2$, 2D $WTe_2$ can be etched to NRs structures (Fig. 5b). The surface energies of the (100) and (010) facets of 1T' $WTe_2$ are calculated to be 0.0315 and 0.0158 eV/Å$^2$, respectively, based on the density functional theory (DFT) calculation. The higher energy of (100) facet than (010) indicates that the [100] direction is the preferred growth direction for 1T' $WTe_2$ NRs (Fig. 5f and Supplementary Note 3). For the 2D layered $WTe_2$, etching always happens at the two ends of the 1T' WTe$_2$ (Supplementary Fig. 4), and thus the prototype of NRs structures can be established. In the third stage, H$_2$ etching and ripening occur concurrently (Fig. 5c). The wide WTe$_2$ can be etched to W and H$_2$Te, and W then migrates into the liquid Te. The tiny crystalline nuclei can grow at the two ends of the 1T' WTe$_2$ and larger NRs will then grow at the cost of smaller ones due to the energy difference according to Ostwald ripening, suggesting a reasonably growth mechanism of WTe$_2$ NRs[37, 38]. Based on the Binary Te-W phase diagram, as the source ratio (Te/W) increases, more W can be dissolved in the liquid Te[39, 40]. As a result, there would be more migrated W atoms in the liquid Te in the high Te/W ratio system, and the WTe$_2$ NRs with increased length-width ratio can be obtained. Even though all the growth mechanisms are discussed under the specific reaction condition, we believe it can be expanded to the whole Te/W/NaCl reaction system. Furthermore, this growth mechanism could also be used to understand many other 2D to 1D TMD synthesis processes.

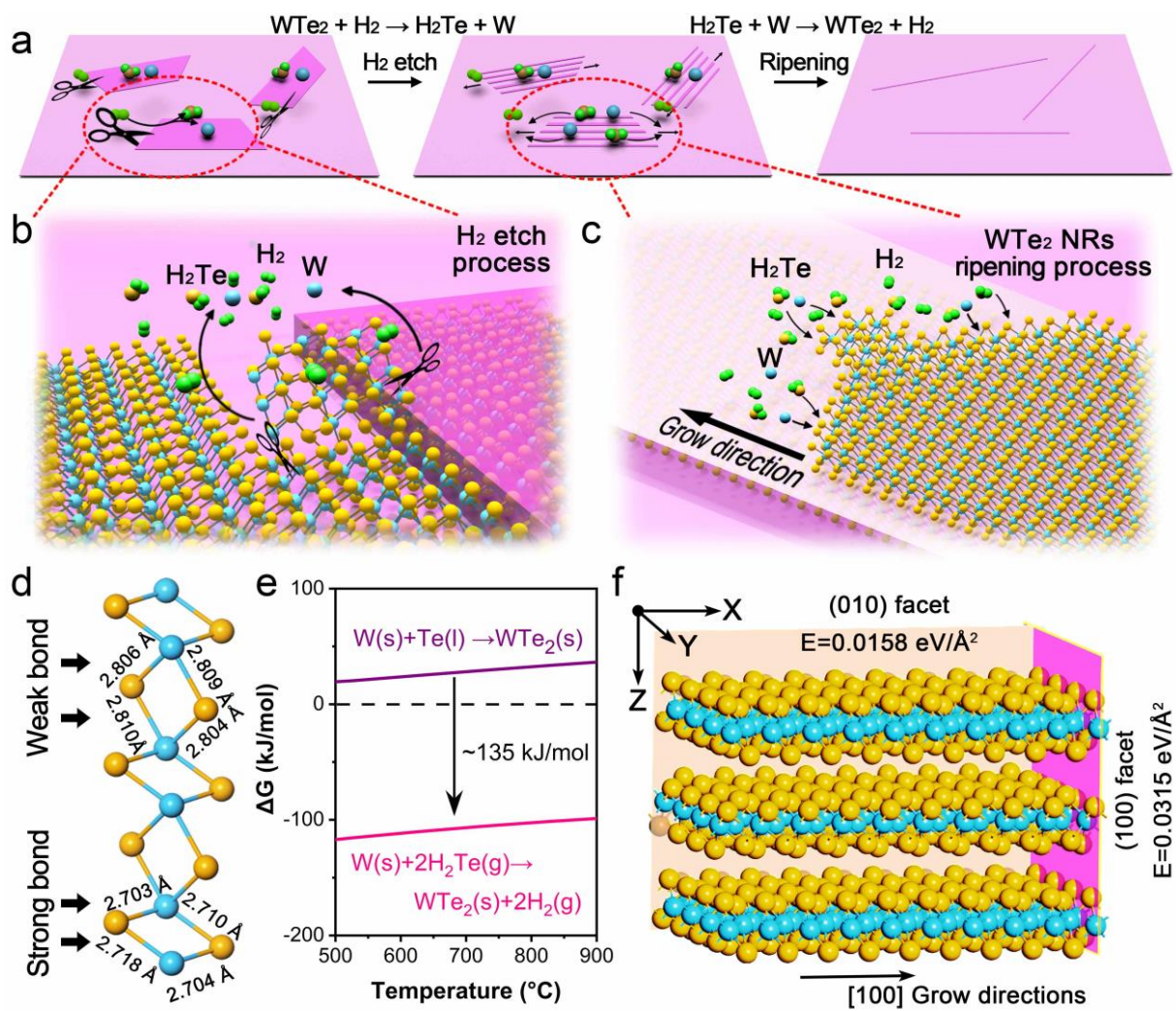

**Fig. 5** Growth mechanism of 1T' WTe$_2$ NRs. **a, b, c** Proposed growth mechanism model of 1T' WTe$_2$ NRs and the corresponding reaction schematics of (**b**) H$_2$ etch process and (**c**) ripening process. **d** Atomic structure of 1T' WTe$_2$ indicating different bond lengths of W-Te. **e** The changes of Gibbs free energy for conversion without and with H$_2$. **f** Crystal structure of 1T' WTe$_2$ with exposed (100) and (010) facets.

**Discussion**

In summary, ML has been successfully applied to guide the CVD synthesis of 1D few-layered 1T' WTe$_2$ NRs. Relevant characterizations have demonstrated the high-quality of as-grown samples. High AUROC of 0.93 indicates the excellent performance of the well-trained XGBoost-C model. Feature importance extracted from the model suggests that $H_2$ gas flow rate and reaction temperature are the most important features governing the synthesis of 1T' $WTe_2$ NRs. Optimization has been further performed to obtain the optimal ranges of each synthesis parameter with high "Can grow" probability. Additionally, the effect of source ratio (Te/W) on the length-width ratio of as-obtained $WTe_2$ NRs has been investigated systematically, and a strong positive correlation is demonstrated. A growth mechanism of $WTe_2$ NRs has further been proposed, where 1D $WTe_2$ were found to be evolved from the 2D $WTe_2$ with the aid of $H_2$. Our work, integrating ML analysis and experimental results, will shed light on the accelerated development of 1D materials and boost the diversified nanostructures derived from 2D TMDs family.

**Methods**

**Synthesis of few-layered $WTe_2$ NRs**. 2 mg powders (Te:W:NaCl=16:1:0.1) were mixed by the agate mortar and then placed on the 10 mm × 30 mm Si/$SiO_2$ substrate (with a 285 nm oxidation layer). Another 10 mm × 30 mm Si/$SiO_2$ substrate was put on top of the powders with the polished surface down, forming a sandwiched structure. They were then transferred in to a closed (inner) quartz tube (with a diameter of 12 mm and length of 300 mm). The (inner) quartz tube was placed in the quartz tube (with a diameter of 2 inches) and then positioned at the center of the furnace. The quartz tube was cleaned with the Ar gas (99.99%) for 20 min before heating. After that, the mixed gas of $H_2$/Ar (40/100 sccm, respectively) was used as the carrier gas. The furnace was heated to 750 °C in 20 min, and then kept for 4 min. Then, the furnace cooled down to room temperature naturally and the WTe$_2$ NRs can be obtained on the substrate.

**Characterization.** The microstructures and morphology of as-grown WTe$_2$ were characterized by optical microscopy (Olympus BX51), Raman (WITEC Alpha 200R), XPS (Kratos Analytical, Manchester), AFM (Asylum Research Cypher S), and STEM (JEOL ARM-200F (S)). The STEM characterization was carried out with CEOS CESCOR aberration corrector, operated at an accelerating voltage of 80 kV. The EDX mapping results were carried out with the TEM. The Raman characterization was carried out with a 532 nm laser (The laser power used was less than 1 mW, the system was calibrated with the Raman peak of Si at 520 cm$^{-1}$). The XPS analysis was performed by a Kratos Axis Supra instrument, and the binding energy correction for charging was performance by assigning a value of 284.8 eV to the adventitious C1 s line.

**Acknowledgements**

The authors gratefully acknowledge financial support from the Singapore National Research Foundation under NRF Award (No. NRF-RF2013-08), MOE under AcRF Tier 2 (MOE2016-T2-1-131), MOE under AcRF Tier 3(MOE2018-T3-1-002), the National Natural Science Foundation of China (Grants 61701402, 61804125, 11904289 and 61974120), the Fundamental Research Funds for the Central Universities (Grants 3102019PY004 and 31020190QD010), the Key Program for International Science and Technology Cooperation Projects of Shaanxi Province (Grant 2018KWZ-08), the Natural Science Foundation of Shaanxi Province (Grant 2019JQ-613), the start-up funds from Northwestern Polytechnical University (Grant 19SH020159 and 19SH020123), and the Northwest University Doctorate Dissertation of Excellence Funds (Grant YYB17020).

**Author Contributions**

M. Xu and B. Tang contributed equally to this work. Z. Zhang, X. Wang and Z. Liu conceived and supervised the project. M. Xu, L. Zheng, X. Wang and Z. Liu designed the experiments. M. Xu prepared and characterized the samples. B. Tang and Y. Lu constructed the ML models and performed the ML result analysis. C. Zhu conducted the STEM characterization. C. Zhu, J. Zhang, and N. Han carried out the data fitting, DFT and Gibbs energy calculation. Y. Guo, J. Di, P. Song, Y. He collected the synthesis data. L. Kang performed the XPS analysis. M. Xu and B. Tang wrote the paper. All authors discussed the results and commented on the manuscript.

**Conflicts of interest**

The authors declare no competing financial interest.